\documentclass[twoside,reqno]{mbook}
\usepackage{epsfig,cite}
\usepackage{amssymb,amsmath}
\usepackage{times}
\begin{document}

\sloppy \raggedbottom

\setcounter{page}{1}

\newpage
\setcounter{figure}{0}
\setcounter{equation}{0}
\setcounter{footnote}{0}
\setcounter{table}{0}
\setcounter{section}{0}
                                                                                                    

\hyphenation{cor-res-pon-ding}
                                                                                                    
\title{Nucleon Structure and Generalized Parton Distributions}
                                                                                                    
\runningheads{Nucleon Structure and Generalized Parton Distributions}{
Eric~Voutier}
                                                                                                    
\begin{start}
\author{Eric Voutier}{}
                                                                                                    
\address{Laboratoire de Physique Subatomique et de Cosmologie\\
IN2P3-CNRS / Universit\'e Joseph Fourier\\
53 avenue des Martyrs\\
38026 Grenoble cedex, France}{}
                                                                                                    
\begin{Abstract}
This paper discusses a selected part of the experimental program dedicated to 
the study of Generalized Parton Distributions, a recently introduced concept 
which provides a comprehensive framework for investigations of the partonic 
structure of the nucleon. Particular emphasis is put on the Deeply Virtual 
Compton Scattering program performed at the Jefferson Laboratory. The short and 
long term future of this program is also discussed in the context of the several
experimental efforts aiming at a complete and exhaustive mapping of Generalized
Parton Distributions.
\end{Abstract}
\end{start}
                                                                                                    
\section{Introduction}
                                                                                                    
From an experimental point of view, the story of the nucleon structure started
in the fifties when deviations from the Mott cross section were observed in
elastic electron scattering, meaning that the nucleon was no longer a pointlike
object~\cite{Hof55}. The size of the nucleon is embedded in the so-called
electromagnetic form factors which characterize the nucleon shape with respect
to the electromagnetic interaction. This shape depends on the resolution of the
probe which is controlled by the momentum transfer $Q^2$ to the nucleon, and
which can also be seen as the size of the volume to which the virtual photon 
couples. This leads to the non-relativistic picture of the form factors as Fourier
transforms of the charge and magnetisation densities of the nucleon.
                                                                                                    
At the end of the sixties, an unexpected result was obtained in Deep Inelastic
electron Scattering (DIS) where it was found that the cross section for
excitation energies well beyond the resonance region was weakly depending on the 
momentum transfer as compared to elastic scattering~\cite{Bre69}. This behaviour 
was later identified as the first evidence of the existence of quarks. The cross 
section for these experiments depends on the additional variable $x_B$, that is 
the fraction of the nucleon longitudinal momentum carried by the quarks, and can 
be expressed in terms of the probability to find in the nucleon a parton of 
given longitudinal momentum. This feature led to extensive measurements of 
momentum and spin distributions of quarks into polarized and unpolarized 
nucleons, the so-called parton distributions whose statisfactory knowledge has 
now been obtained after thirty years of experimental efforts.
                                                                                                    
However, the puzzle of the spin structure of the nucleon is not yet resolved.
DIS experiments determined the contribution of the quarks spin to the nucleon
spin to amount only to 20-30~\% of the total spin~\cite{Ash88}. This surprising
result suggests that the gluons spin may play a significant role in this
problem. The recent results of the COMPASS experiment do not support this
hypothesis~\cite{Age06}. What is the origin of the nucleon spin and what is
the importance of orbital momentum are still unanswered questions. The
Generalized Parton Distribution (GPD) framework can contribute to this problem
within a comprehensive picture that unifies form factors, parton distributions,
and the total angular momentum of quarks.
                                                                                                    
The next section introduces the general concept of GPDs and some remarkable
properties that link GPDs to the usual observables of the nucleon structure. 
The experimental access to GPDs is then discussed in the context of the Deeply
Virtual Compton Scattering (DVCS) process. Recent results from Jefferson
Laboratory (JLab) experiments are further presented before addressing briefly
the forthcoming experimental programs at the different lepton facilities.
                                                                                                    
\section{Generalized Parton Distributions}
                                                                                                    
\begin{figure}[h]
\centerline{\epsfig{file=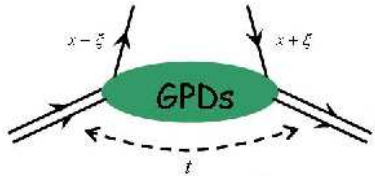,width=50mm}}
\caption{Symbolic representation of GPDs.
\label{GPD-diag}}
\end{figure}
                                                                                                    
GPDs are four universal functions ($H^q$, $E^q$, ${\widetilde H}^q$, and 
${\widetilde E}^q$ where $q$ denotes the quark flavor) describing the 
non-perturbative quark structure of the nucleon (correspondingly four gluon 
GPDs describe the gluon structure)~\cite{{Mul94},{Rad97},{Ji97}}. They 
correspond to the overlap probability $\Psi^{\ast}(x-\xi) \cdot \Psi(x+\xi)$ 
of picking a quark in the nucleon and inserting it back with a different (or 
same) spin, longitudinal momentum, and transverse position 
(fig.~\ref{GPD-diag}). GPDs depend on three parameters: $x$ the initial 
longitudinal momentum of the quark, $\xi$ the transferred longitudinal momentum 
or skewness parameter, and $t$ the momentum transfer to the nucleon that can be 
interpreted as the Fourier conjugate of the transverse position in the impact 
parameter space~\cite{Bur00}. The richness of the GPDs parametrization of the 
nucleon sub-structure is in their non-diagonal feature which, among others, 
allows for initial nucleon spin-flip, a source of information which is not 
accessible with DIS. A simple physics 
picture~\cite{{Bur00},{Ral02},{Die02},{Bel02}} has been proposed which
allows to interpret GPDs as the $1/Q$ resolution distribution in the transverse 
plane of partons with longitudinal momentum $x$, constituting a 
femto-tomography of the nucleon.
                                                                                                    
The optical theorem provides a link between the forward limit of the Compton 
scattering amplitude and DIS, leading to the relations between GPDs and parton 
distributions
\begin{equation}
H^q(x, \xi=0, t=0) = q(x) \ \ \ \ {\widetilde H}^q(x, \xi=0, t=0) = \Delta q(x) 
\ \ .
\end{equation}
The first moment of each GPD identifies to a distinct form factor of the 
nucleon:
\begin{eqnarray}
\int_{-1}^{+1} dx \, H^q(x,\xi,t) = F_1^q(t) & \ \ \ \ & \int_{-1}^{+1} dx \, 
{\widetilde H}^q(x,\xi,t) = g_A^q(t) \\
\int_{-1}^{+1} dx \, E^q(x,\xi,t) = F_2^q(t) & \ \ \ \ & \int_{-1}^{+1} dx \, 
{\widetilde E}^q(x,\xi,t) = g_P^q(t)
\end{eqnarray}
where the $\xi$ independence results from Lorentz invariance. The second moment 
of $H$ and $E$ GPDs are linked together within Ji'sum rule~\cite{Ji97} to the 
total angular momentum of quarks:
\begin{equation}
J^q = \frac{1}{2} \Delta \Sigma + L^q = \frac{1}{2} \int_{-1}^{+1} dx \, \, x 
\left[ H^q(x,\xi,t=0) + E^q(x, \xi, t=0) \right] \ \ .
\label{Jisumrule}
\end{equation}
This last expression has generated a lot of interest: considering that DIS 
provides the spin part of  the angular momentum, the knowledge of $H$ and $E$ 
GPDs allows to access the quark orbital momentum. 

\section{Deeply Virtual Compton Scattering}
 
GPDs can experimentally be accessed in exclusive leptoproduction of photons and 
mesons. The latter case, where the GPD information is convoluted with a 
distribution amplitude describing the out-going meson, and which requires the 
selection of longitudinal virtual photons is not discussed here. A comprehensive 
review can be found in~\cite{Die03}.
 
\begin{figure}[t]
\centerline{\epsfig{file=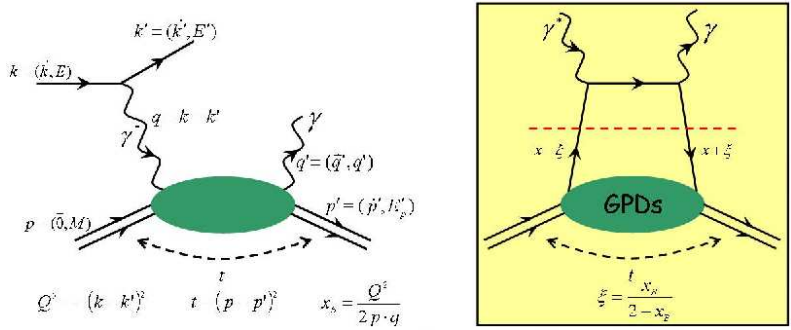,width=105mm}}
\caption{DVCS representation: kinematic variables (left), and handbag diagram
(right). \label{DVCS-Hand}}
\end{figure}

DVCS, corresponding to the absorption of a virtual photon by a nucleon followed 
quasi-instantaneously by the emission of a real photon (fig.~\ref{DVCS-Hand}), 
is the simplest reaction which allows access to GPDs. This process became very 
important for nucleon structure studies when it was shown that, in the Bjorken 
limit, the leading contribution to the reaction amplitude could be represented 
by the so-called handbag diagram (fig.~\ref{DVCS-Hand}). The important feature 
of this representation is the factorization~\cite{{Ji98},{Col99}} of the 
reaction amplitude in a known hard part corresponding to the photon-quark 
interaction and an unknown soft part related to GPDs.
 
In addition to the DVCS amplitude, the cross section for leptoproduction of 
photons gets contributions from the Bethe-Heitler (BH) process where the real 
photon is emitted by the initial or final lepton, leading to
\begin{equation}
\frac{d^5 \sigma}{dQ^2 dx_B dt d\phi_e d \varphi} = {\cal T}_{BH}^2 + {\vert 
{\cal T}_{DVCS} \vert}^2 + 2 \, {\cal T}_{BH} \Re e\{ 
{\cal T}_{DVCS} \}
\end{equation}
where $\varphi$ is the out-of-plane angle between the leptonic and hadronic 
planes. Though the BH amplitude dominates the cross section at JLab energies, 
it is a completely known process exactly calculable from the nucleon 
electromagnetic form factors. Polarization degrees of freedom help to overcome 
this problem from their sensitivity to the interference between the BH and DVCS 
amplitudes. For instance, the polarized cross section difference for opposite 
beam helicities can be expressed as~\cite{{Gui98},{Ben06}}
\begin{eqnarray}
\frac{d^5 \Delta \sigma}{dQ^2 dx_B dt d\phi_e d \varphi} & = & \frac{1}{2} 
\left[ \frac{d^5 \overrightarrow{\sigma}}{dQ^2 dx_B dt d\phi_e d \varphi} - 
\frac{d^5 \overleftarrow{\sigma}}{dQ^2 dx_B dt d\phi_e d \varphi} \right] \\
& = & {\cal T}_{BH} \, \Im m \{{\cal T}_{DVCS}\} + \Re e \{{\cal T}_{DVCS}\} \,
\Im m \{{\cal T}_{DVCS}\} \nonumber
\end{eqnarray}
where the imaginary part of the DVCS amplitude appears now linearly instead of 
quadratically, and magnified by the BH amplitude. These observables can be 
decomposed in terms of harmonics with respect to $\varphi$~\cite{BelM02} 
leading, in the twist-3 approximation, to
\begin{eqnarray}
\frac{d^5 \Delta \sigma}{dQ^2 dx_B dt d\phi_e d \varphi} & = & \Gamma_2(x_B,
Q^2,t) \, s_1^{DVCS} \sin(\varphi) \\
& + & \frac{\Gamma_3(x_B,Q^2,t)}{P_1(\varphi)P_2(\varphi)} \, \left[ s_1^I 
\sin(\varphi) + s_2^I \sin(2\varphi) \right] \nonumber
\end{eqnarray}
where $\Gamma_i$ are kinematical factors and $P_i$ are the BH propagators. In 
this expression, $s_1^{DVCS}$ and $s_2^I$ are twist-3 coefficients while $s_1^I 
= k \, \Im m\{ C^I({\cal F})\}$ is a twist-2 coefficient directly linked to the 
linear combination of GPDs
\begin{equation}
C^I({\cal F}) = F_1 {\cal H} + \frac{x_B}{2-x_B}\, (F_1+F_2) \widetilde{\cal H} 
- \frac{t}{4M^2} F_2 {\cal E}
\label{GPD:comb}
\end{equation}
with for example
\begin{eqnarray}
{\cal H} & = & \sum_q {\cal P} \int_{-1}^{+1} dx \, \left( \frac{1}{x-\xi} + 
\frac{1}{x+\xi} \right) H^q(x,\xi,t) \nonumber \\
& - & i \pi \sum_q e_q^2 \left[ H^q(\xi,\xi,t) - H^q(-\xi,\xi,t) \right]
\end{eqnarray}
$e_q$ being the electric charge of the considered quark in unit of the 
elementary charge. The dominance of a twist-2 contribution to the cross section 
is a strong indication for factorization and enables a GPD based interpretation.
 
\section{Recent Results from Jefferson Laboratory}
 
The first DVCS candidate signal was reported by the H1 
collaboration~\cite{Adl01} from a deviation observed in the photon 
electroproduction cross section as compared to the BH cross section: in the H1 
energy range, the DVCS process dominates the cross section. At smaller energy, 
the interference between the DVCS and the BH amplitudes was observed 
successively at HERMES~\cite{Air01} and CLAS~\cite{Ste01} as a characteristic 
$\sin(\varphi)$ dependence of the relative beam spin asymmetry. These last 
experiments are strong evidences of the existence of a DVCS signal but do not 
tell about the reliability of a GPD based interpretation whose prerequisite is 
an experimental proof of the factorization of the cross section. This motivates 
in part the experimental program at JLab.
 
\subsection{The E00-110 $p({\vec e},e' \gamma)p$ and E03-106 $n({\vec e},e'
\gamma)n$ Hall A Experiments}
 
The E00-110~\cite{E00-110} and E03-106~\cite{E03-106} experiments have been
taking data successively on hydrogen and deuterium in the hall A of JLab,
investigating different issues of the DVCS process: in the former case, the
test of handbag dominance in the valence region ($x_B$=~0.36) between $Q^2
$=~1.9~GeV$^2$ and 2.3~GeV$^2$ is under concern while the latter measurement is
an exploratory attempt to access $E$, the least known and constrained GPD which
directly enters Ji's sum rule (eq.~\ref{Jisumrule}).
 
\begin{figure}[t]
\centerline{\epsfig{file=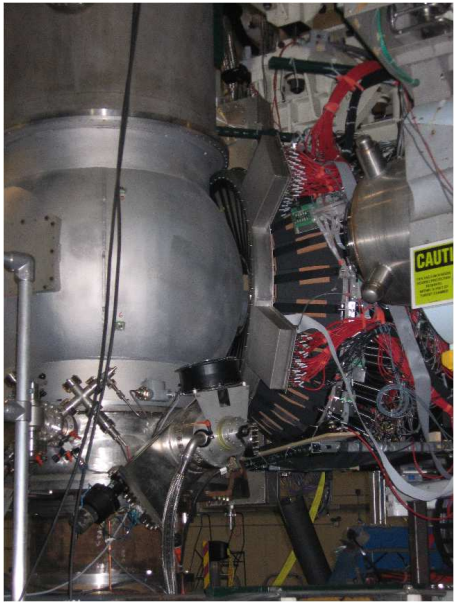,width=60mm}}
\caption{The E00-110 and E03-106 experimental setup in the Hall A of JLab.
\label{Dvcs_detn}}
\end{figure}

A specific experimental setup (fig.~\ref{Dvcs_detn}) has been instrumented
which involves a new reaction chamber (RC), a PbF$_2$ electromagnetic
calorimeter (EC), a recoil detector (RD), and customized electronics and data
acquisition~\cite{Cam05}: the new RC is optimized in thickness and features a
larger exit beam pipe, reducing the otherwise very high rate of M\o ller
electrons; the {\v C}erenkov sensitive material of the EC insures the rapidity
of the delivered signal and a reduced sensivity to hadronic background; the RD
allows to check the exclusivity of the reaction selection; the read-out
electronics based on Analog Ring Samplers~\cite{Lac00} resolves the pile-up of
signals separated from at least 5~ns. These many features allow the detector
operation in the highly hostile environment of an electromagnetic facility:
current luminosities of 4$\times$10$^{37}$~cm$^{-2} \cdot$~s$^{-1}$ were 
achieved during deuterium data taking.
 
\begin{figure}[t]
\centerline{\epsfig{file=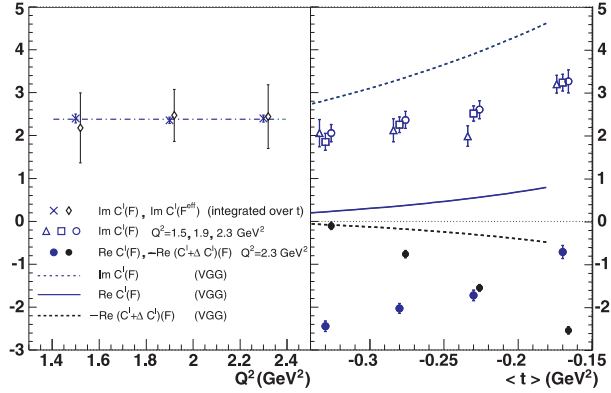,width=80mm}}
\caption{$Q^2$ and $t$ dependences of the harmonic coefficients extracted from
E00-110 measurements~\cite{Mun06}; curves labelled VGG are twist-2 calculations 
neglecting the contribution of the GPD $E$~\cite{{Van99},{Goe01},{Gui05}}.
\label{fig5:prl}}
\end{figure}

From the measurement of beam helicity dependent cross sections for photon
electroproduction, the $\varphi$ harmonic structure of the sum and the
difference of polarized cross sections for opposite beam helicities (Sec. 3) is
investigated neglecting contributions of the squared DVCS
amplitude~\cite{Mun05}. Experimental results for the proton case~\cite{Mun06}
are reported on fig.~\ref{fig5:prl}: the twist-2 ($\Im m\{ C^I({\cal F})\}$)
and twist-3 ($\Im m\{ C^I({\cal F}^{eff})\}$) harmonics are
essentially $Q^2$ independent; thanks to different kinematical factors the
twist-3 contribution to the cross section turns out to be small; the general
trend of the $t$ dependence of the different harmonics is consistent with a GPD
based calculation while the exact magnitude is not reproduced. These features
are evidences of the factorization of the cross section at $Q^2$ as small as
2~GeV$^2$ and support the prediction of perturbative Quantum Chromo-Dynamics
scaling in DVCS~\cite{{Rad97},{Ji97}}. This legitimates a GPD based
interpretation and this experiment provides for the first time a
model-independent measurement of linear combinations of GPDs and GPDs
integrals.
 
Because of the cancellation between the $u$ and $d$ quarks in $\widetilde{H}$ 
and following $F_1(t) \approx 0$ at small $t$, the DVCS process on a neutron 
target turns out to be a unique tool to access $E$ (eq.~\ref{GPD:comb}). The 
$n({\vec e},e' \gamma)n$ polarized cross sections are deduced via the 
subtraction of the proton yield measured with $p({\vec e},e' \gamma)p$ from the 
D$({\vec e},e' \gamma)X$ yield where the residual system $X$ can be either a 
nucleon or a deuteron~\cite{Maz06}. The deconvolution of these two contributions 
is insured by their dynamical separation $\Delta M_X^2 = t/2$ in the missing 
mass spectra. It should be also noticed that the coherent deuterium channel is 
expected to be small and rapidly decreasing with $t$ following the 
electromagnetic form factors dependence. These features allow for a reliable 
extraction of the neutron DVCS cross section~\cite{Maz06-Th}.
 
\subsection{The E01-113/E06-003 $p({\vec e},e' \gamma)$ Hall B Experiment}
 
Similar measurements with an unpolarized hydrogen target~\cite{E01-113} have
been recently performed in hall B with the aim of studying GPDs from both 
electroproduction of photons and mesons. Thanks to the acceptance of the
CLAS detector~\cite{Mec03}, a large phase space in $(Q^2,x_B,t)$ is explored in
terms of relative beam spin asymmetries (BSA) and cross sections.
 
This first dedicated DVCS experiment in Hall B required additions to CLAS: an
electromagnetic calorimeter consisting of 424 PbW0$_4$ cristals is installed in
the central part of CLAS for the $\gamma$ detection between 4 and 16 degrees;
in order to allow operation of this device, a superconducting solenoidal magnet
placed prior to the calorimeter confines low energy M\o ller electrons in the
beam pipe~\cite{Gir06}. The selection of the DVCS process is insured by the  
triple coincidence detection of the scattered electron, the recoil proton, 
and the produced real photon. 

It should be noticed that while BSA are in principle easier to measure than 
cross sections, their GPD interpretation is basically more difficult. The 
E00-110 experiment shows that in the JLab energy range, the BSA $\varphi$ 
structure is more complex than the simple $\sin(\varphi)$ assumed in earlier 
experiments. Consequently, the harmonic coefficients can only be extracted 
through iterative procedures involving some GPD parametrizations, meaning that 
BSA interpretation in terms of GPDs is model dependent.  

\subsection{The E05-114 ${\vec p}({\vec e},e' \gamma p)$ Hall B Experiment}

\begin{figure}[h]
\centerline{\epsfig{file=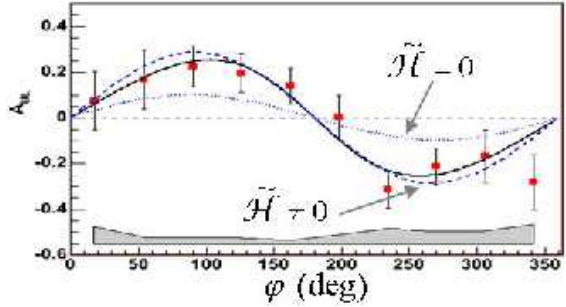,width=75mm}}
\caption{Target spin asymmetry measured by the CLAS collaboration on a 
longitudinally polarized hydrogen target~\cite{Che06}. The dashed and dotted 
curves are model predictions showing the sensitivity to $\widetilde{\cal{H}}$, 
and the solid curve is a fit to the data neglecting the $\varphi$ dependence of 
the denominator.     
\label{AUL}}
\end{figure}
 
Beam polarization observables are not unique tools in DVCS to reveal the GPD 
content of the nucleon. There exists indeed a complete set of observables
involving beam and/or target polarization: each observable gives access to a
different linear combination of GPDs, and at least four different observables
are required to unravel GPDs. For instance, the difference between polarized 
cross sections for opposite longitudinal polarization of a proton target
measures the combination 
\begin{eqnarray}
C^I_{LP}({\cal F}) & = & \frac{x_B}{2-x_B} ( F_1 + F_2 ) \left(  {\cal H} + 
\frac{x_B}{2} {\cal E} \right) \label{GPD:AUL} \\
& &  + \, F_1 \widetilde{\cal H} - \frac{x_B}{2-x_B} \, \left( \frac{x_B}{2} 
F_1 + \frac{t}{4M^2} F_2 \right) \widetilde{\cal E} \nonumber
\end{eqnarray}
where $C^I_{LP}({\cal F})$ plays an equivalent role to $C^I({\cal F})$ of 
eq.~\ref{GPD:comb}. Because of the kinematical factors weighting each GPD,
the combination of eq.~\ref{GPD:AUL} is expected to be sensitive to 
$\widetilde{H}$. This feature was observed in an experiment performed in the
Hall B of JLab where the target spin asymmetry (TSA) of the DVCS process was 
measured for a longitudinally polarized hydrogen target~\cite{Che06}. The
results reported on fig.~\ref{AUL} show that a GPD based model taking into
account the contribution of $\widetilde{H}$ is more likely able to reproduce 
the data than the same calculation neglecting this contribution. This
observation motivates the E05-114 dedicated experiment~\cite{E05-114} which 
will investigate the TSA and the double spin asymmetry (DSA) with the CLAS 
detector over a large phase space in the variables $(Q^2,x_B,t)$: the TSA will
measure $\Im m\{ C^I_{LP}({\cal F})\}$ while the DSA involving beam and target
polarizations will measure $\Re e\{ C^I_{LP}({\cal F})\}$.

\section{Perspectives}

\begin{figure}[h]
\centerline{\epsfig{file=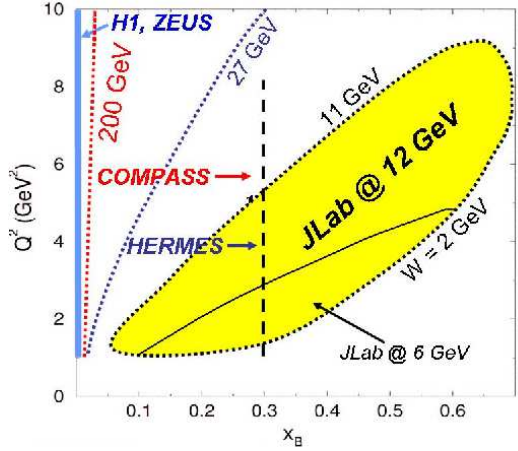,width=70mm}}
\caption{Experimental phase space relative to the different programs 
investigating GPDs. \label{Jlab12}}
\end{figure}
  
The fundamental character of the GPD framework motivates several experimental 
programs at major lepton facilities whose combination provides a systematic
mapping of these distributions (fig.~\ref{Jlab12}). Depending on the available 
beam energy, different aspects of the nucleon structure can be investigated: H1 
and ZEUS experiments at DESY probe the GPDs in the very small $x_B$ region
relevant for the gluon content of the nucleon, HERMES~\cite{Now06} at DESY and
COMPASS~\cite{{COM05},{Bur06}} at CERN extend their investigation up to the 
valence quark region, and the energy upgrade of 
JLab~\cite{{PR12-06-114},{PR12-06-119}} allows to access the high $x_B$ domain. 
The remarkable complementarity between these different experiments will provide 
a comprehensive picture of the nucleon structure, including the flavor 
decomposition of GPDs which is achieved from deeply virtual meson production 
and/or proton and neutron DVCS.  
  
\section{Conclusions}

The study of nucleon structure is living very exciting moments. The
factorization of the DVCS cross section on the proton was established at JLab, 
opening acces to GPDs. A worldwide very promising experimental investigation of 
the GPD framework is starting. In a near future, the dedicated program at JLab 
and COMPASS will deliver unprecedented information on the quark and gluon 
content of the nucleon, hopefully unraveling the nucleon spin puzzle and the 
quark confinement. 

\section*{Acknowledgments}

I would like to thank the organizers of the XXVth International Workshop on
Nuclear Theory for their invitation and warm hospitality at Rila Mountains.
  
This work was supported in part by the U.S. Department of Energy (DOE) contract
DOE-AC05-06OR23177 under which the Jefferson Science Associates, LLC, operates
the Thomas Jefferson National Accelerator Facility, the National Science
Foundation, the French Atomic Energy Commission and National Center of
Scientific Research.

\end{document}